\documentclass[prl,aps,twocolumn,groupedaddress,floats,showpacs,final]{revtex4-1}
\usepackage{graphicx}
\usepackage{dcolumn}
\usepackage{bm}
\usepackage{color}
\usepackage{ulem}
\usepackage[caption=false]{subfig}
\definecolor{DarkGreen}{rgb}{0,0.7,0.08} 
\definecolor{Grey}{rgb}{0.5,0.5,0.5}
\definecolor{Red}{rgb}{0.8,0.3,0.3}
\definecolor{Green}{rgb}{0.1,0.8,0.1}
\definecolor{Blue}{rgb}{0.1,0.1,0.8}
\newcommand{\oldtext}[1]{}

\begin{document}

\title{Ground state phase diagram of parahydrogen in one dimension}
\author{Massimo Boninsegni}
\affiliation{Department of Physics, University of Alberta, Edmonton, Alberta, Canada, T6G 2E1}
%\affiliation{$^2$ {Department of Physics, University of Massachusetts, Amherst, MA}}
\date{\today}                                           % Activate to display a given date or no date

%-------------------------------------------------------------------------------------------------------------
\begin{abstract}
The low temperature phase diagram of parahydrogen in one dimension is studied by Quantum Monte Carlo simulations, whose results are  interpreted within the framework of Luttinger liquid theory. We show that, contrary to what claimed in a previous study  (Phys. Rev. Lett. {\bf 85}, 2348 (2000)), the equilibrium phase is a crystal. The phase diagram mimics that of parahydrogen in two dimensions, with a single quasi-crystaline phase and no quantum phase transition, i.e., it is qualitatively  different from that of $^4$He in one dimension.

\end{abstract}

%Uncomment for PACS numbers title message
%\pacs{00.00, 20.00, 42.10}
% Keywords required only for MST, PB, PMB, PM, JOA, JOB? 
%\vspace{2pc}
%\noindent{\it Keywords}: Article preparation, IOP journals
% Uncomment for Submitted to journal title message
%\submitto{\JPA}
% Comment out if separate title page not required
\maketitle
The physics of strongly interacting many-body systems confined to one physical dimension (1D) is a 
subject of longstanding theoretical interest, early works dating back to the 1960s. A number of exact mathematical statements can be made about 1D quantum fluids \cite{mattis}, whose relevance is not merely ``academic" as there exist ways to restrict the motion of small atoms or molecules to 1D, to a  degree enabling one  to test experimentally 
some of the most interesting predictions. For example, one can adsorb gases made of small atoms or molecules, such as helium, inside carbon nanotubes \cite{teizer}, or in porous glasses \cite{beamish}, providing cylindrical confinment for the adsorbed atoms that can approach the 1D limit, as the radius of the cylinders becomes of the order of a nanometer or less \cite{nanotubes}. 
This has motivated theoretical studies of hard core fluids such as $^4$He \cite{mille,saverio} and parahydrogen ({\it p}-H$_2$) \cite{cazzoni} in strictly 1D, as well as in models of  confinements aimed at describing reasonably realistically the environment experienced by atoms and molecules moving inside a single nanotube \cite{kazzoni}, or in the interstitial channel of a bundle of nanotubes \cite{crespi}, or in nanopores \cite{delmaestro}, as well as  inside the (quasi-1D) core of a screw dislocation in solid $^4$He \cite{pollet}. 
\\ \indent
What makes 1D many-body systems particularly interesting, is the existence of an elegant and powerful 
formalism, known as Luttinger Liquid Theory (LLT), which provides a universal description of 
Bose or Fermi systems in terms of linear quantum hydrodynamics \cite{llt}. The LLT asserts that, while no true long range order can exist in 1D, of either crystalline or superfluid type (even at temperature $T=0$),  two-body correlations display a slow power-law decay as a function of distance $r$. Consider for definiteness the pair correlation function $g(r)$; at any finite temperature $T$, it will take on the following behavior in the thermodynamic limit:
\begin{equation}\label{maineq}
g(r)\sim 1-\frac {1}{2\pi^2 K\rho^2 r^2}+A\  {\rm cos}(2\pi\rho r)\frac {1} {\rho^{2} r^{2/K}}
\end{equation}
Here, $\rho$ is the linear density of particles, $A$ a non-universal, system-dependent constant, whereas the Luttinger parameter $K$ determines how fast the oscillation of the $g(r)$ decays at long distances, and can thus be used to draw a meaningful distinction between ``quasi-crystalline" and ``quasi-superfluid" phases.
Specifically, if $K > 2$ the static structure factor will develop (Bragg) peaks at reciprocal lattice vectors, which is the experimental signature of a crystalline solid. On the other hand, if $K < 2$ the system possesses no density quasi long-range order, and can be  regarded either as a a glassy insulator or a ``superfluid" (i.e., featuring off-diagonal quasi long range order), with a different degree of  robustness against disorder or  external potentials, depending on the  value of $K$ (a more precise classification is not necessary for the purpose of this work, as will be shown below). 
Henceforth, we shall be using the adjectives ``superfluid" and ``crystalline" in this sense.
\\ \indent
The zero temperature phase diagram of $^4$He in 1D has been extensively studied theoretically. The equilibrium phase is a very low density, weakly self-bound superfluid, which transitions to a crystalline phase at higher density \cite{saverio}. Of at least equal interest is the physics of {\it p}-H$_2$ in 1D, for several reasons, both fundamental as well as practical. Parahydrogen molecules are spin-zero bosons of mass half of that of $^4$He, leading to the speculation that {\it p}-H$_2$ could be a potential superfluid at low $T$. There is strong experimental and theoretical evidence suggesting that, while small cluster of {\it p}-H$_2$ may indeed display superfluid properties at low $T$ \cite{sindzingre,noi,noi2,toennies}, no fluid phase of {\it p}-H$_2$ can be stabilized at $T=0$, neither in three nor in two dimensions \cite{me04}. Rather, the only thermodynamically stable phase is a (non-superfluid) crystal, due to the strength of the attractive well of the intermolecular potential. It has been claimed, however, that in 1D {\it p}-H$_2$ might be a Luttinger superfluid \cite{cazzoni}. This would be potentially important a result, as one could conceivably explore ways of achieving three-dimensional superflow of {\it p}-H$_2$  through a network of interconnected quasi-1D channels \cite{pollet}, for example in a porous glass such as vycor. 
It should also be noted that the study of {\it p}-H$_2$ in reduced dimensions has also potential technological revelance, given the current interest in the storage of hydrogen in nanostructures, e.g., carbon nanotubes, for fueling purposes \cite{storage}.
\\ \indent
In this Letter, we study the zero temperature phase diagram of {\it p}-H$_2$ in 1D, by means of first principle Quantum Monte Carlo simulations at finite temperature. We first extrapolate computed energy estimates to $T=0$, thereby obtaining the equation of state, and then characterize the phases of the system as a function of linear density through the long-range behavior of the pair correlation function $g(r)$, which is directly accessible by simulation, without uncontrolled approximations.
\\ \indent
The main result of this study is that the phase diagram of  {\it p}-H$_2$ in 1D mimics that of the system in two and three dimensions. The equilibrium phase is a crystal, i.e., a phase featuring slowly decaying oscillations in the pair correlation function, with a value of the Luttinger parameter $K\sim 3.5$, increasing at higher density. The system possesses quasi-long-range order even at negative pressure, i.e., below the equilibrium density, all the way down to the spinodal. No evidence of any other phase is observed (nor, obviously, of any quantum phase transition). Thus, the physics of the system is qualitatively different from that of $^4$He in 1D.
This is in striking disaccord with what claimed in the only previous study of {\it p}-H$_2$ in 1D (Ref. \cite{cazzoni}). We attribute the incorrect physical conclusion reached therein, to the approach utilized, focusing on the energy rather than on the quantities of relevance, namely correlation functions.
\\ \indent
We model our system of interest as a collection of $N$ {\it p}-$H_2$ molecules, regarded as point-like, moving in a 1D region of length $L$, with periodic boundary conditions. The many-body Hamiltonian is the following:
\begin{equation}\label{ham}
\hat H = -\lambda\sum_i\frac{\partial^2}{\partial x_i^2}+\sum_{i<j} V(r_{ij})
\end{equation}
where $\lambda=12.031$ K\AA$^2$ and $r_{ij}\equiv |x_i-x_j|$. The linear density is $\rho=N/L$. For consistency with previous calculations, we choose the well-known Silvera-Goldman potential \cite{silvera}  to model the interaction $V$ between a pair of {\it p}-H$_2$ molecules. 
\\ We computed low-temperature thermodynamics for the system described by Eq. (\ref{ham}) by means of Quantum Monte Carlo simulations at finite temperature, based on the Worm Algorithm in the continuous-space path integral  representation. This methodology allows one to compute numerically exact estimates of thermodynamic quantities for Bose systems at finite $T$.  Its most important quality in this work is that it grants access to the pair correlation function, which can be obtained in an unbiased way.
Details of the simulation are standard \cite{worm,worm2,cuervo}.  
We have carried out simulations of a system described by the above microscopic Hamiltonian, comprising $N=30, 60$ and 120 particles, respectively. 
\begin{figure}[h]
\includegraphics[scale=0.39]{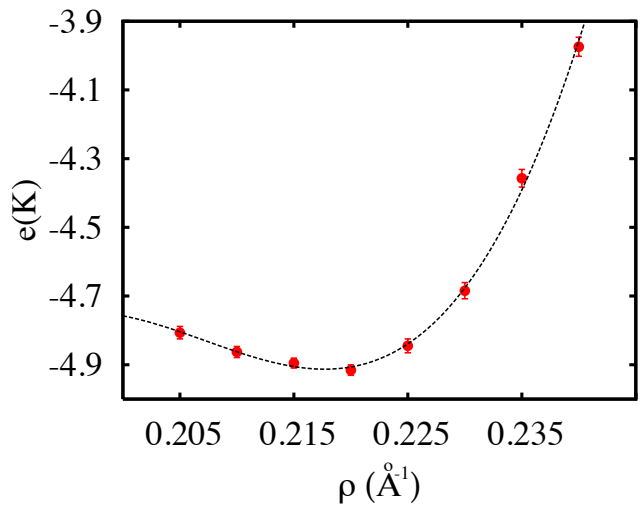}
\caption{(Color online)  Ground state energy per  molecule (in K) computed by QMC for 1D {\it p}-H$_2$ as a function of the linear density $\rho$ (in \AA$^{-1}$).  Dashed line is a polynomial fit  to the data. When not shown, statistical errors are smaller than the size of the symbols.}
\label{f1}
\end{figure}
\begin{table}[h]
\begin{tabular}{ccc} \hline\hline
$\rho$&This work  &Ref. \onlinecite{cazzoni}  \\ \hline
0.220  &$-4.916(15)$ &-4.834(7)  \\
0.240&$-3.974(28)$ &  \\
0.265&1.661(21) &  \\
0.290  &18.37(7) &19.203(10) \\ 
0.329 &94.37(4) &97.963(16) \\ \hline\hline 
\end{tabular}\hfil\break 
\caption {\noindent {Ground state energy per molecule (in K) for 1D {\it p}-H$_2$ at various linear densities $\rho$ (in \AA$^{-1}$), extrapolated to the thermodynamic limit. Also shown for comparison are ground state estimates from Ref. \onlinecite{cazzoni}. Statistical errors, in parentheses, are on the last digit(s).}
}\label{tableone}
\end{table} 
\\ \indent
We begin by discussing the low temperature energetics, specifically the $T$=0 equation of state of 1D {\it p}-H$_2$. All energy results reported here are extrapolated to the thermodynamic ($N\to\infty$) limit; no difference is seen between energy estimates obtained for $N$=60 and $N$=120, within statistical uncertainties. Moreover, results shown correspond to a temperature $T=0.25$ K, but they should be regarded as ground state estimates, as 
thermal averages remain unchanged below $T=0.5$ K.
Fig. \ref{f1} shows the energy per molecule computed as a function of the linear density $\rho$;  Table \ref{tableone} reports a few representative values, comparing them to the corresponding estimates from Ref. \cite{cazzoni}, where a ground state study of the system was carried out based on Diffusion Monte Carlo (DMC) simulations with the same Hamiltonian and pair potential.
A polynomial fit to the data yields an equilibrium density $\rho_e=0.2178\pm0.007$ \AA$^{-1}$ and a spinodal density $\rho_s = 0. 209\pm 0.001$ \AA$^{-1}$. Although our ground state energy estimates are consistently lower than those of Ref. \cite{cazzoni} (by small but not insignificant amounts), our estimates of the equilibrium and spinodal densities  are  in agreement with theirs.  The disagreement lies in the characterization of the phase(s) of the system, as we show below by means of an examination of the pair correlation function. 
\begin{figure}[h]
\includegraphics[scale=0.39]{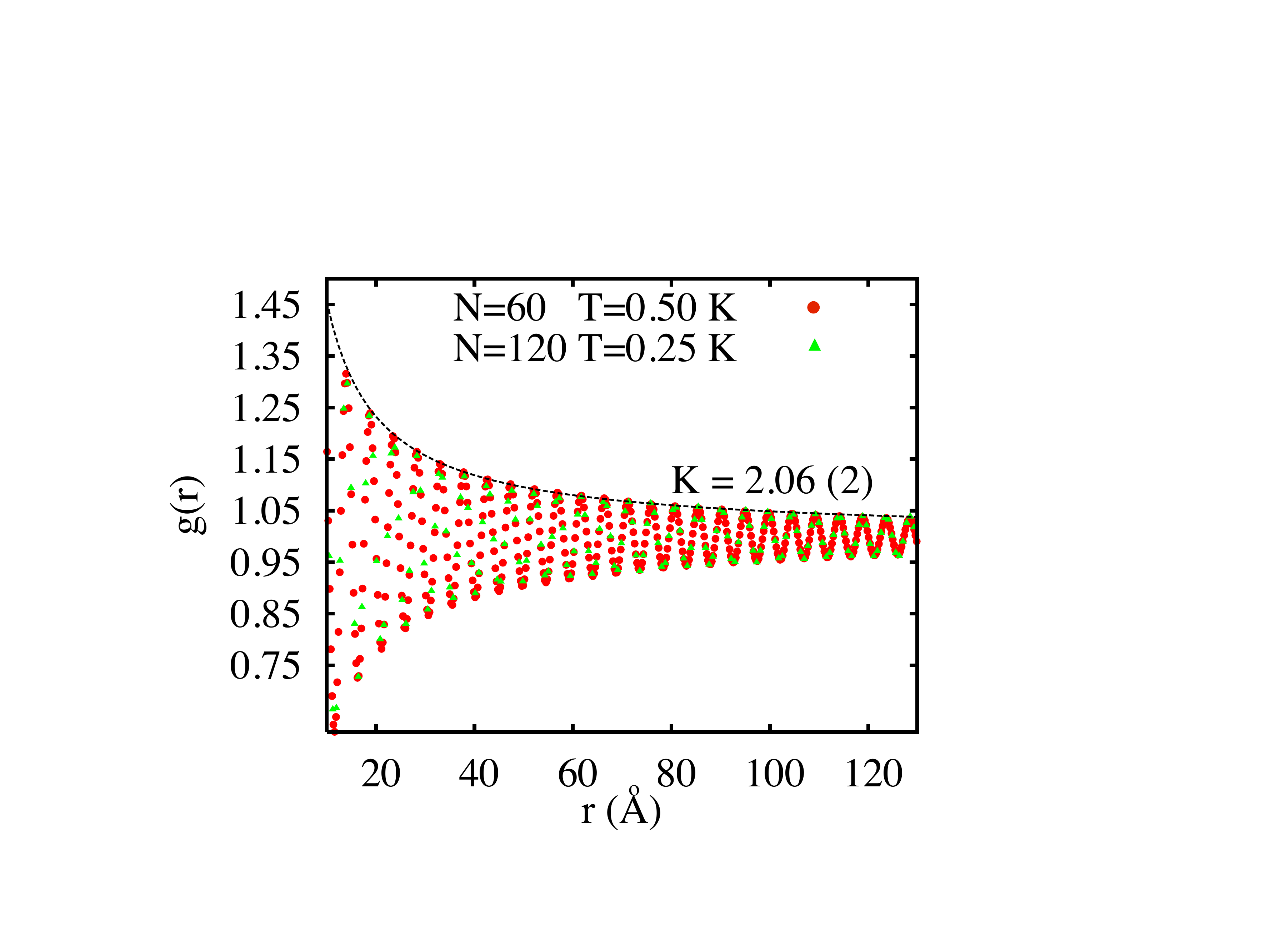}
\caption{(Color online)  Pair correlation function $g(r)$ for 1D {\it p}-H$_2$ at a linear density $\rho=0.21$ \AA$^{-1}$. Two sets of data are displayed, one pertaining to a simulation at temperature $T=0.5$ K, for a system comprising $N=60$ particles (circles), and at $T=0.25$ K and $N=120$ (triangles). Statistical errors are smaller than symbols. Dashed line is a fit to the maxima of the functions obtained with the expression 
$f(r)=1+A/r^{2/K}$, with the Luttinger parameter $K=2.06(2)$.}
\label{f2}
\end{figure}

Figure \ref{f2} shows the pair correlation function $g(r)$, computed at low temperature  for 1D {\it p}-H$_2$ at a linear density $\rho=0.210$ \AA$^{-1}$, i.e., well below the equilibrium density, at the lower edge of the region of metastability. According to the LLT, in the $T\to 0, \ L\to\infty$ limit, the physical properties of the system only depend on the product $LT$. Data shown in Fig. \ref{f2} pertain to two physical system, one comprising $N=60$ particles at a temperature $T=0.5$ K, the other with $N=120$ and at $T=0.25$ K. Within the statistical uncertainties of the calculations, the data fall on the same curve, as expected. We have  observed this scaling behaviour at all linear densities considered here, at these two temperatures.\\
In order to determine the value of the Luttinger parameter $K$, we can fit the data at large $r$ to the
expression (\ref{maineq}), or, which is easier, fit the maxima of the oscillating function to 
$f(r)=1+A/{r^{2/K}}$, with $A$ and $K$ fitting parameters. In this case, the procedure yields $K=2.06\pm0.02$, i.e., a value only barely above 2, but already in the crystalline sector, as described by LLT.
\\ \indent
On increasing the linear density, the physics does not change qualitatively;  $K$ increases monotonically, i.e., the crystalline nature of the ground state is enhanced as the system is compressed. 
\begin{figure}[h]
\includegraphics[scale=0.39]{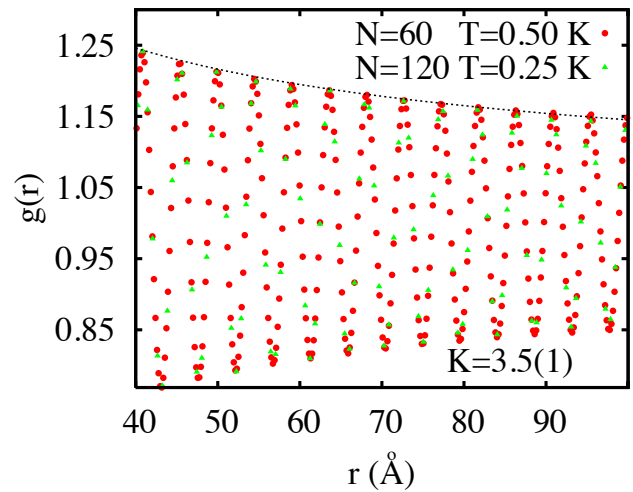}
\caption{(Color online)  Pair correlation function $g(r)$ for 1D {\it p}-H$_2$ at a linear density $\rho=0.22$ \AA$^{-1}$. Two sets of data are displayed, one pertaining to a simulation at temperature $T=0.5$ K, for a system comprising $N=60$ particles (circles), and at $T=0.25$ K and $N=120$ (triangles). Statistical errors are smaller than symbols. Dashed line is a fit to the maxima of the functions obtained with the expression 
$f(r)=1+A/r^{2/K}$, with the Luttinger parameter $K=3.5(1)$.}
\label{f3}
\end{figure}
For example, close to the equilibrium density, at $\rho=0.22$ \AA$^{-1}$, it is $K=3.5\pm0.1$, i.e., 
well in the crystalline side of the LLT. Results for this case are displayed in Figure \ref{f3}. Although, as the linear density increases, obtaining an accurate estimate of $K$ becomes more difficult (because the decay of the oscillation is slower), nonetheless the expected monotonic increase of $K$ is readily established.  For example, $K\sim 10$ at $\rho=0.265$ \AA$^{-1}$.
\\ \indent
There is therefore no evidence of any quantum phase transition in the entire domain of existence of a thermodynamically (meta)stable uniform phase. Indeed, within the known and well understood 
differences that the reduction of dimensionality entails, the ground state phase diagram of 1D {\it p}-H$_2$ is qualitatively identical with that in 2D and 3D, i.e., only a crystalline phase is present. This physical conclusion constitutes another strong piece of evidence that prospects of observing superfluid behaviour in {\it p}-H$_2$ (or anything that could be meaningfully regarded as such) are dim, as the tendency to crystallize remains strong, even when the system is maximally confined.
Such a phase diagram is  very different from the case of $^4$He, whose equilibrium phase in 1D is indeed a Luttinger superfluid, and which undergoes a quantum phase transition to a crystal at high density. 
\\ \indent
As mentioned in the introduction,  the first calculation of the equation of state of {\it p}-H$_2$ in 1D was carried out in Ref. \cite{cazzoni}. The statement was made therein that the equilibrium phase is a liquid, and that a quantum phase transition between a liquid and a crystal  takes place at a density $\bar\rho$ close to  0.312 \AA$^{-1}$. In contrast, no evidence whatsoever of any phase transition is seen in this work, and certainly nowhere near $\bar\rho$, at which the system is in fact a ``hard" crystal.  A (quasi) crystalline phase is the only one observed in this work, at and below the equilibrium density, all the way to the spinodal (below which it is actually observed in the simulation to break down into crystal clusters).
Obviously, it is necessary to examine this disagreement in detail and identify its origin, as both the calculations carried out in Ref. \cite{cazzoni} and here are based on the same microscopic Hamiltonian, and employ numerical techniques  which should yield compatible results, within statistical errors. As we argue below, the disagreement arises not from differences in the numerical data, which exist but are relatively unimportant, but rather from  their physical interpretation.
\\ \indent
The physical conclusion of Ref. \cite{cazzoni} was reached by comparing ground state 
energy estimates yielded by a DMC calculation in which two different initial trial wave function were utilized, a ``fluid-like", enjoying translational invariance, and a ``solid-like", in which {\it p}-H$_2$ molecules are pinned at lattice positions. On carrying out DMC projections based on these two {\it ansatze}, it was found that the energy estimate arrived at by starting from a ``liquid-like" initial ansatz was lower (higher) than that obtained starting from the ``solid-like" trial wave function for $\rho < \bar\rho$ ($\rho > \bar\rho$). This was interpreted as evidence of a liquid-like ground state for $\rho < \bar \rho$, and crystalline for $\rho > \bar \rho$.
\\ \indent
A few comments are in order here. The first is that {{the determination of  phase boundaries through a comparison of ground state energy estimates yielded by DMC using different trial wave functions, might be
conceptually justified when the various {\it ansatze} are orthogonal to each other, reflecting the different 
physical pictures that they describe. Its use for a Bose system is puzzling, especially in 1D where no sharp 
distinction can be drawn between liquid and crystalline phases. The ground state wave function of a Bose system
is positive-definite, and therefore in principle a DMC projection starting from {\it any}  positive-definite trial many-body wave function (as both {\it ansatze} utilized in Ref. \cite{cazzoni} are)  should converge to the same value, given a sufficiently long  projection time. Different energy expectation values obtained with different {{non-negative}} trial wave functions in a {{finite }}  projection time, are { {far more}} likely to be the consequence of a poorer relative optimization of one wave function with respect to the other, as well as of sampling and/or 
population size bias issues \cite{bm}, { {than to have}} a genuine physical origin. 
\\ \indent
Looking at the data in Table \ref{tableone} one can see that the ground state 
estimates of Ref. \cite{cazzoni} are consistently above the ones obtained here; at the highest densities, the difference between the value reported here and that of Ref. \cite{cazzoni} is an order of magnitude greater than that between the estimates yielded by the  DMC projection with the two different {\it ansatze} 
(Table I of Ref. \cite{cazzoni}), a fact clearly pointing to failure to achieve convergence 
to the true ground state  with {\it either} starting trial wave function. 
\\ \indent
More importantly, the energy is not the relevant quantity at which to look, in order to assess the physical nature of a many-body state. Rather, as mentioned in the introduction  the long-range behavior of the pair correlation function (considered here) or the off-diagonal one-body density matrix, contain in 1D all the information required to characterize the system. The long-distance behaviour of neither one of these quantities was examined at all in Ref. \cite{cazzoni}, ostensibly on the assumption that it would largely mimic that built into the trial wave function.  There is in principle no reason, however, why the projection algorithm should not build the kind of long-range, slowly decaying correlations that are observed here, even if they are missing in the initial trail wave function. Only
on the basis of such correlations can any statement be made about the  nature of the ground state. 
A limitation of DMC, of course, is that unbiased expectation values of quantities other than the energy are generally difficult to obtain \cite{hinde}. Indeed, empirical evidence is mounting that finite temperature techniques are a better option to investigate the ground state of Bose systems.
\\
\indent
Summarizing, we have computed the ground state equation of state of {\it p}-H$_2$ in 1D, and found that the phase diagram of the system mimics qualitatively that in 2D and 3D, namely only phase is present. Whereas in 2D and 3D such a phase is a crystal in a strict sense, in 1D it possesses  crystalline quasi long-range order as defined in the context of Luttinger Theory. No evidence of any superfluid phase, again as defined within LLT is seen, including metastable ones.
\\ \indent
This work was supported in part by the Natural Science and Engineering Research Council of Canada. MB gratefully acknowledges conversations with Nikolay Prokof'ev and Boris Svistunov, as well as 
hospitality of the University of Massachusetts, Amherst, where part of this research was conducted.

\end{document}